\documentclass[a4paper,twocolumn]{esapub} 
\usepackage{natbib,graphicx,epsfig}

\title{The mechanism of Swing Absorption of fast magnetosonic waves
in inhomogeneous media}
\author{B.M. Shergelashvili$^{(1),}$\thanks{On leave from the Center
for Plasma Astrophysics, Abastumani Astrophysical Observatory,
Kazbegi Ave. 2a, Tbilisi 380060, Georgia}, T.V.
Zaqarashvili$^{(2)}$, S. Poedts$^{(1)}$ and B. Roberts$^{(3)}$}

\affil{$^{(1)}$Centre for Plasma Astrophysics, Katholieke
 Universiteit Leuven, Celestijnenlaan 200B, B-3001
Leuven, Belgium, E-mails:Bidzina.Shergelashvili@wis.kuleuven.ac.be
and Stefaan.Poedts@wis.kuleuven.ac.be\\
$^{(2)}$Abastumani Astrophysical Observatory, Al. Kazbegi ave. 2a,
380060
Tbilisi, Georgia, E-mail: temuri@hubble.uib.es\\
$^{(3)}$School of Mathematics and Statistics, University of St.
Andrews, St. Andrews, Fife, KY16 9SS, Scotland, UK, E-mail:
bernie@mcs.st-and.ac.uk}

\begin{document}


\maketitle

\begin{abstract}

The recently suggested swing interaction between fast magnetosonic
and Alfv\'en waves  \citep{paper1} is generalized to inhomogeneous
media. We show that the fast magnetosonic waves propagating across
an applied non-uniform magnetic field can parametrically amplify
the Alfv\'en waves propagating along the field through the
periodical variation of the Alfv\'en speed. The resonant Alfv\'en
waves have half the frequency and the perpendicular velocity
polarization of the fast waves. The wavelengths of the resonant
waves have different values across the magnetic field, due to the
inhomogeneity in the Alfv\'en speed. Therefore, if the medium is
bounded along the magnetic field, then the harmonics of the
Alfv\'en waves, which satisfy the condition for onset of a
standing pattern, have stronger growth rates. In these regions the
fast magnetosonic waves can be strongly {\lq absorbed\rq}, their
energy going in transversal Alfv\'en waves. We refer to this
phenomenon as {\lq {\it Swing Absorption}\rq}. This mechanism can
be of importance in various astrophysical situations.

\end{abstract}

\section{Introduction}

Wave motions play an important role in many astrophysical
phenomena. Magnetohydrodynamic (MHD) waves may transport momentum
and energy, resulting in heating and acceleration of an ambient
plasma. A variety of waves have recently been detected in the
solar atmosphere using the SOHO and TRACE spacecraft. Hence, an
understanding of the basic physical mechanisms of excitation,
damping and the interaction between the different kinds of MHD
wave modes is of increasing interest \citep{rob1}. Formally
speaking, there is a group of direct mechanisms of wave excitation
by external forces (e.g.\ turbulent convection, explosive events
in stellar atmospheres, etc.) and wave dissipation due to
non-adiabatic processes in a medium (such as viscosity, thermal
conduction, magnetic resistivity, etc.). There is a separate group
of wave amplification and damping processes due to resonant
mechanisms \citep{goos,poed}. This means that particular wave
modes may be damped (amplified) due to energy transfer
(extraction) into or from other kinds of oscillatory motions, even
when wave dissipation is excluded from consideration.

Recently, a new kind of interaction between different MHD wave
modes, based on a parametric action, has been suggested
\citep{zaq,paper1,zaq0,zaq2,zaq3}. In this case the mechanism of
wave interaction originates from a basic physical phenomenon known
in classical mechanics as {\lq parametric resonance\rq}, occurring
when an external force (or oscillation) amplifies the oscillation
through a periodical variation of the system's parameters. The
mechanical analogy of this phenomenon is a mathematical pendulum
with periodically varying length. When the frequency of the length
variation is twice the frequency of the pendulum oscillation, the
amplitude of the oscillation grows exponentially in time. Such a
mechanical system can consist of a pendulum (transversal
oscillations) with a spring (longitudinal oscillations). A
detailed description of such a mechanical system is given in
\citet{paper1} (hereinafter referred to as Paper I).

Here we present a brief report on our recent results on swing
interactions of fast and Alfv\'en modes in inhomogeneous media
(Shergelashvili et al. (2004)).

\section{Basic equations and equilibrium model}
\begin{figure}
\centering
\includegraphics[width=0.99\linewidth]{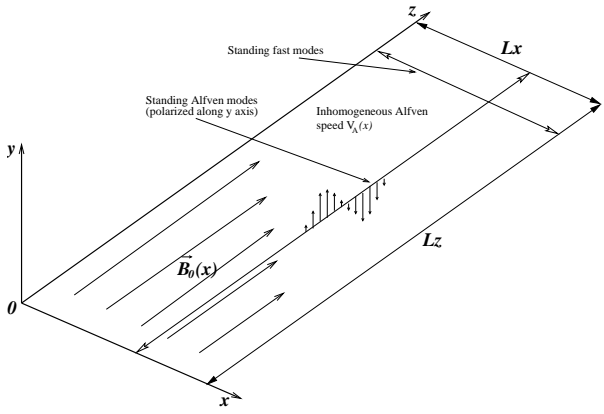}
\caption{Schematic view of the inhomogeneous background
configuration and the directions of wave propagation
(polarization).}
\end{figure}

Consider a magnetized medium with zero viscosity and infinite
conductivity, where processes are assumed to be adiabatic. Then
the macroscopic dynamical behaviour of this medium is governed by
the ideal magnetohydrodynamic (MHD) equations:
\begin{equation}\label{eqcont}
\frac{\partial \rho }{\partial t}+\vec{\nabla}\mathbf{\cdot }(\rho
\vec{U}%
)=0,
\end{equation}%
\begin{equation}\label{eqmot}
\rho \frac{\partial \vec{U}}{\partial t}+\rho \left( \vec{U}\cdot
\vec{\nabla%
}\right) \vec{U}=-\vec{\nabla}\left[ p+\frac{B^{2}}{8\pi }\right]
+\frac{%
\left( \vec{B}\cdot \vec{\nabla}\right) \vec{B}}{4\pi },
\end{equation}%
\begin{equation}\label{eqind}
\frac{\partial \vec{B}}{\partial t}+\left( \vec{U}\cdot
\vec{\nabla}\right) \vec{B}=\left( \vec{B}\cdot
\vec{\nabla}\right) \vec{U}-\vec{B}\left(
\vec{%
\nabla}\cdot \vec{U}\right) ,
\end{equation}%
\begin{equation}\label{eqener}
\frac{\partial p}{\partial t}+\left( \vec{U}\cdot
\vec{\nabla}\right)
p=%
\frac{\gamma p}{\rho }\left( \frac{\partial \rho }{\partial
t}+\left(
\vec{U}%
\cdot \vec{\nabla}\right) \rho \right) ,
\end{equation}
where $p$ and $\rho$ are the plasma pressure and density, $\vec U$
is the velocity, $\vec B$ is the magnetic field of strength
$B=\vert B \vert$ and $\gamma$ denotes the ratio of specific
heats.

We consider an equilibrium magnetic field directed along the $z$
axis of a Cartesian coordinate system,
$\vec{B}_{0}=(0,0,B_{0}(x))$. The equilibrium magnetic field
$\vec{B}_{0}$ and density $\rho _{0}$ are inhomogeneous in $x$.
The force balance condition, from Eq.~(\ref{eqmot}), gives the
total (thermal + magnetic) pressure in the equilibrium to be a
constant:
\begin{equation}\label{balance}
p_{0}(x)+\frac{B^2_{0}(x)}{8\pi } = {\rm constant}.
\end{equation}
Also, the equation of state
\begin{equation}
p_{0}(x)= p_0(T_0(x),\rho_0(x)),
\end{equation}
relates the unperturbed pressure, density $\rho_0(x)$ and
temperature $T_0(x)$.

Eqs.~(1)-(4) are linearized around the static ($\vec{U_0}=0$)
equilibrium state (5). This enables the study of the linear
dynamics of magnetosonic and Alfv\'en waves. A schematic view of
the equilibrium configuration is shown in Figure 1.
\section{Fast magnetosonic waves}
Now let us consider the propagation of '{\lq pure\rq} fast
magnetosonic waves across the magnetic flux surfaces, i.e.\ along
the $x$ axis, taking {\bf $\vec u=(u_x,0,0)$} and $\partial/
\partial y \equiv 0$, $\partial/
\partial z \equiv 0$. The equation governing the dynamics of fast magnetosonic waves
propagating across the magnetic field lines in the inhomogeneous
medium can be obtained from equations
(\ref{eqcont})-(\ref{eqener}) as \citep{rob1981,rob3}:
\begin{equation}\label{eqmain}
\frac{d }{d x}\left[ \left( \gamma p_{0}(x) +  \frac{%
B^2_{0}(x)}{4\pi }\right) \frac{%
d u(x)}{d x} \right] + \omega^2 \rho _{0}(x)u(x)=0.
\end{equation}
where $u_{x}$ is the velocity perturbation.

The solution of this equation for different particular equilibrium
conditions can be obtained either analytically or numerically.
Equation (\ref{eqmain}) describes the propagation of a fast wave
with speed $(C_s ^2+V_A ^2)^{1/2}$, where $C_s=(\gamma p_0/\rho
_0)^{1/2}$ is the sound speed and $V_A=(B_0 ^2/4 \pi \rho
_0)^{1/2}$ is the Alfv\'en speed.

We study Eq. (\ref{eqmain}) in accordance with the boundary
conditions:
\begin{equation}\label{bounf}
u(0)=u(L_x)=0,
\end{equation}
corresponding to fast waves bounded by walls located at $x=0$ and
$x=L_x$. These boundary conditions make the spectrum of fast modes
discrete: Eq.~(\ref{eqmain}) has a nontrivial solution only for a
discrete set of frequencies,
\begin{equation}\label{freq}
\omega=\omega _n =\omega _0, \, \omega _1,\ldots.
\end{equation}
In this case the solutions for different physical quantities can
be represented as
\begin{displaymath}
u_{x}=\alpha v(x)\sin (\omega _n t), \hskip 0.2cm \rho _{1}=\alpha
r(x)\cos (\omega _n t),
\end{displaymath}
\begin{equation}
b_{z}=\alpha h(x)\cos (\omega _n t),
\end{equation}
where the density and magnetic field perturbations are related to
the velocity perturbations through
\begin{equation}
r(x)=\frac{1}{\omega _n}\left( v(x)\frac{d\rho _{0}}{dx}+\rho
_{0}\frac{dv(x)%
}{dx}\right),
\end{equation}
\begin{equation}
h(x)=\frac{1}{\omega _n}\left(
v(x)\frac{dB_{0}}{dx}+B_{0}\frac{dv(x)}{dx} \right).
\end{equation}
Here (and subsequently) we use a subscript $n$ to denote the
frequency of a given standing fast mode.
\section{Swing amplification of Alfv\'en waves}
Now consider Alfv\'en waves that are linearly polarized in the $y$
direction and propagate along the magnetic field (see Figure 1).
In the linear limit these waves are decoupled from the
magnetosonic waves and the equation governing their dynamics is:
\begin{equation}\label{alflin}
\frac{\partial ^2 b_{y}}{\partial t^2}-V_A ^2 (x)\frac{\partial ^2
b_{y}}{\partial z^2}=0,
\end{equation}
where $V_A(x)$ is Alfv\'en speed. It is clear from
Eq.~(\ref{alflin}) that the phase speed of this mode depends on
$x$ parametrically. Therefore, an Alfv\'en wave with a given wave
length propagates with a '{\lq local\rq} characteristic frequency.
Each magnetic flux surface can evolve independently in this
perturbation mode.

\subsection{Propagating Alfv\'en waves}
Let us now address the non-linear action of the fast magnetosonic
waves, considered in the previous section, on Alfv\'en waves. We
study the weakly non-linear regime. This means that the amplitudes
of the fast magnetosonic waves are considered to be large enough
to produce significant variations of the environment parameters,
which can be felt by propagating Alfv\'en modes, but too small to
affect the Alfv\'en modes themselves. Hence, the magnetic flux
surfaces can still evolve independently. Therefore, as in paper~I,
the non-linear terms in the equations arising from the advective
derivatives $u_{x}\partial b_{y}/\partial x$ and $\left( \rho
_{0}+\rho _{1}\right) u_{x}\partial u_{y}/\partial x$ are assumed
to be negligible. Under these circumstances the governing set of
equations takes the form (see Paper I):
\begin{equation}\label{alfnon1}
\frac{\partial b_{y}}{\partial t}=\left( B_{0}+b_{z}\right)
\frac{\partial u_{y}}{\partial z}-\frac{\partial u_{x}}{\partial
x}b_{y},
\end{equation}%
\begin{equation}\label{alfnon2}
\left( \rho _{0}+\rho _{1}\right) \frac{\partial u_{y}}{\partial
t}=\frac{%
B_{0}+b_{z}}{4\pi }\frac{\partial b_{y}}{\partial z}.
\end{equation}
These equations describe the parametric influence of fast
magnetosonic waves propagating across the magnetic field on
Alfv\'en waves propagating along the field. An analytical solution
of Eqs. (\ref{alfnon1}) and (\ref{alfnon2}) is possible for a
standing pattern of fast magnetosonic waves, the medium being
assumed bounded in the $x$ direction.

Combining Eqs.~(\ref{alfnon1}) and (\ref{alfnon2}) we obtain the
following second order partial differential equation:
\begin{displaymath}
\frac{\partial ^{2}b_{y}}{\partial t^{2}}+\left[ \frac{\partial
u_{x}}{%
\partial x}-\frac{1}{B_{0}+b_{z}}\frac{\partial b_{z}}{\partial
t}\right] \frac{\partial b_{y}}{\partial t} +
\end{displaymath}
\begin{equation}
+\left( \frac{\partial ^{2}u_{x}}{\partial t\partial
x}-\frac{1}{B_{0}+b_{z}}\frac{\partial b_{z}}{\partial
t}\frac{%
\partial u_{x}}{\partial x}\right) b_{y}-\frac{(B_{0}+b_{z})^{2}}{4\pi
\left( \rho _{0}+\rho _{1}\right) }\frac{\partial
^{2}b_{y}}{\partial
z^{2}}%
=0.
\end{equation}
Writing
\begin{equation}
b_{y}=h_{y}(z,t)\exp \left[ -\frac{1}{2} \int \left(
\frac{\partial u_{x}}{\partial
x}-\frac{1}{B_{0}+b_{z}}\frac{\partial b_{z}}{\partial t}\right)
dt\right],
\end{equation}
we obtain
\begin{equation}\label{eqalvin}
\frac{\partial ^{2}h_{y}}{\partial t^{2}}+\frac{1}{2}\left[ S_1
(x,t)-S_2 (x,t)\right ]h_{y}-S_3 (x,t)\frac{\partial
^{2}h_{y}}{\partial z^{2}}=0,
\end{equation}
where,
\begin{equation}\label{s1}
 S_1  = \frac{\partial
^{2}u_{x}}{\partial t\partial
x}+\frac{1}{B_{0}+b_{z}}\frac{\partial ^{2}b_{z}}{\partial t^{2}},
\end{equation}
\begin{equation}\label{s2}
S_2= \frac{1}{\left ( B_0+b_z\right ) ^2}\left (\frac{\partial
b_z}{\partial t}\right ) ^2+\frac{1}{2}\left [ \frac{\partial
u_{x}}{\partial x}+\frac{1}{B_{0}+b_{z}}\frac{\partial
b_{z}}{\partial t}\right ]^2,
\end{equation}
\begin{equation}\label{s3}
 S_3 = \frac{(B_{0}+b_{z})^{2}}{4\pi \left( \rho _{0}+\rho
 _{1}\right)}.
\end{equation}
Finally, applying a Fourier analysis with respect to the $z$
coordinate,
\begin{equation}
h_{y}(z,t)=\int \hat{h}_{y}(k_{z},t)e^{ik_{z}z}dk_{z},
\end{equation}
and neglecting the second and higher order terms in $\alpha$, we
obtain the following Mathieu-type equation:
\begin{equation}\label{alfmain}
\frac{\partial ^{2}\hat{h}_{y}}{\partial
t^{2}}+k_{z}^{2}V_{A}^{2}\left[ 1+ \alpha F(x)\cos (\omega _n
t)\right] \hat{h}_{y}=0,
\end{equation}
where
\begin{equation}\label{mathie}
F(x)=2\frac{h(x)}{B_{0}}-\frac{r(x)}{\rho _{0}}-v(x)\frac{ \omega
_n}{2k_{z}^{2}V_{A}^{2}}\frac{1}{B_{0}}\frac{dB_{0}}{dx}.
\end{equation}
It should be noted that the expression (\ref{s2}) for $S_2 (x,t)$
consists only of terms of second order and higher in $\alpha$, and
so can be neglected directly for the case of weakly non-linear
action addressed here.

Equation~(\ref{alfmain}) has a resonant solution when the
frequency of the Alfv\'en mode $\omega _A$ is half of $\omega _n$:
\begin{equation}\label{rescond}
\omega _{A} = k_{z}V_{A}(x)\approx \frac{1}{2}\omega _n.
\end{equation}
This solution can be expressed as
\begin{equation}
{\hat h_y} (k_z,t)={\hat h_y}(k_z,t=0)e^{{{\left |{\delta}\right
|}\over {2\omega_n}}t}\left
   [{\cos}{{\omega_n}\over 2}t - {\sin}{{\omega_n}\over 2}t \right ],
\end{equation}
where
\begin{equation}\label{delta}
\delta (x)=\alpha k_{z}^2 V_{A} ^2(x)F(x).
\end{equation}
The solution has a resonant nature within the frequency interval
\begin{equation}\label{resineq}
\left | \omega _A - \frac{\omega _n}{2} \right |<\left |
\frac{\delta}{\omega _n} \right |.
\end{equation}
Similar expressions have been obtained in the Paper~I for a
homogeneous medium. In that case, the Alfv\'en speed is constant
and, therefore, the fast magnetosonic waves amplify the Alfv\'en
waves with the same wavelength everywhere. In the case of an
inhomogeneous Alfv\'en speed, the resonance condition
(\ref{rescond}) implies that the wavelength of the resonant
harmonics of the Alfv\'en waves depends on $x$. This means that
the fast magnetosonic waves now amplify Alfv\'en waves with
different wavelengths (but with the same frequency) in different
magnetic flux surfaces (i.e.,\ different $x$-values).
\subsection{Standing Alfv\'en waves}
When we consider a system that is bounded in the $z$ direction,
the boundary conditions along the $z$ axis introduce an additional
quantization of the wave parameters. In particular, in this case
each spatial harmonic of the Alfv\'en mode can be represented as
\begin{equation}
\hat{h}_{y} ^{m}=\hat{h}_{y}(k_{m},t)\cos (k_{m}z),
\end{equation}
where $k_{m}=\pi m/L_{z}$ ($m=1,2,\ldots$) and $L_{z}$ is the
characteristic length of the system in the $z$ direction. This
then leads to a further localization of the spatial region where
the swing transfer of wave energy from longitudinal to transversal
oscillations is permitted. The resonant condition (\ref{rescond})
implies that
\begin{equation}\label{rescondm}
k_{m}V_{A}(x_{n,m})\approx \frac{1}{2}\omega _{n}.
\end{equation}
Therefore, the resonant areas are concentrated around the points
$x_{n,m}$ for which condition (\ref{rescondm}) holds. Within these
resonant areas the longitudinal oscillations damp effectively and
their energy is transferred to transversal oscillations with wave
numbers $k_z=k_{m}$ satisfying the resonant conditions. These
resonant areas are localized in space and can be referred to as
regions of {\it swing absorption of the fast magnetosonic
oscillations} of the system. The particular feature of this
process is that the energy transfer of fast magnetosonic waves to
Alfv\'en waves occurs at half the frequency of the fast waves.
\begin{figure*}
\centering
\includegraphics[scale=0.8]{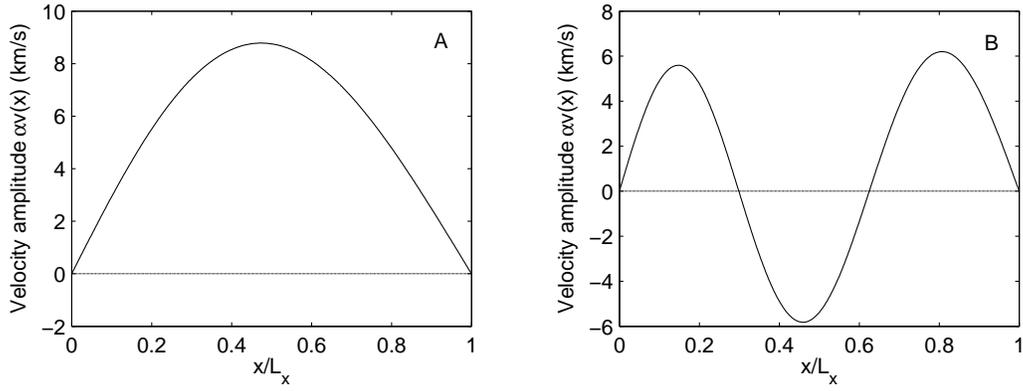}
\caption{Sample solutions of the standing fast magnetosonic modes.
Panel A: zeroth-order harmonic $n=0$, period $T=5.6069$ min.,
$\alpha=0.015$ ; Panel B: second-order harmonic $n=2$, period
$T=1.8514$ min., $\alpha=0.03$}\label{fapa1}.
\end{figure*}

\section{Numerical simulation}

In this section we consider in detail the process of swing
absorption of fast waves into Alfv\'en waves. We consider a
numerical study of equation (\ref{eqmain}) subject to the boundary
conditions (\ref{bounf}). We study, as an example, the case of a
polytropic plasma when both the thermal and magnetic pressures are
linear functions of the $x$ coordinate:
\begin{equation}\label{pprof}
p_{0}=p_{00}+p_{01}\frac{x}{L_x},
\end{equation}
\begin{equation}\label{dprof}
\rho_{0}=C^2\left ( p_{00}+p_{01}\frac{x}{L_x}\right
)^{\frac{1}{\gamma}},
\end{equation}
\begin{equation}\label{bprof}
B_{0}=\sqrt{h_{00}+h_{01}\frac{x}{L_x}},
\end{equation}
where, $p_{00}$, $p_{01}$, $h_{00}$, $h_{01}$ and $C$ are
constants, and $L_{x}$ denotes the length of the system along the
$x$ direction. The pressure balance condition (\ref{balance})
immediately yields
\begin{equation}
p_{01}=-\frac{h_{01}}{8\pi}.
\end{equation}
The solution of the wave equation depends on the values of the
above set of constant parameters. In general, different
equilibrium regimes can be considered including those
corresponding to different limits of the plasma ${\beta}$: $\beta
\ll 1$, $\beta \approx 1$ and $\beta \gg 1$.
\begin{table}[h]
\begin{center}
\begin{tabular}{cccc}
\hline \hline $p_{00}$ dyn/cm$^2$&$p_{01}$ dyn/cm$^2$&$h_{00} $
G$^2$&$h_{01}$ G$^2$\\ \hline
100&-70&$10^{4}$&$1.7593 \cdot 10^{3}$\\
\hline \hline
$L_x$ km & $L_z$ km & $C$ &$\gamma$\\
\hline
15000&$6.5\cdot L_x$&$10^{-6}$&$5/3$\\
\hline \hline
\end{tabular}
\end{center}
\caption{Values of the constant parameters used in the calculation
of our illustrative solutions. The dimension of $C$ is
g$^{1/2}$cm$^{(2-3\gamma)/2\gamma}$/$dyn
^{1/2\gamma}$}\label{tab1}.
\end{table}
The values of all constant parameters are given in
Table~\ref{tab1}. We took arbitrary values of parameters, but they
are somewhat appropriate to the magnetically dominated solar
atmosphere (say the chromospheric network). In Figure~\ref{fapa1}
we show the profiles of $\alpha v(x)$ for the standing wave
solutions, for two cases with different modal `wavelength'. Panels
A and B, respectively, correspond to the characteristic
frequencies: ~$\omega _0\approx 1.87\cdot 10^{-2}$ s$^{-1}$
(period 5.61 min) and ~$\omega_2 \approx 5.66\cdot 10^{-2}$
s$^{-1}$ (period 1.85 min).

\begin{figure*}
\centering
\includegraphics[scale=0.6]{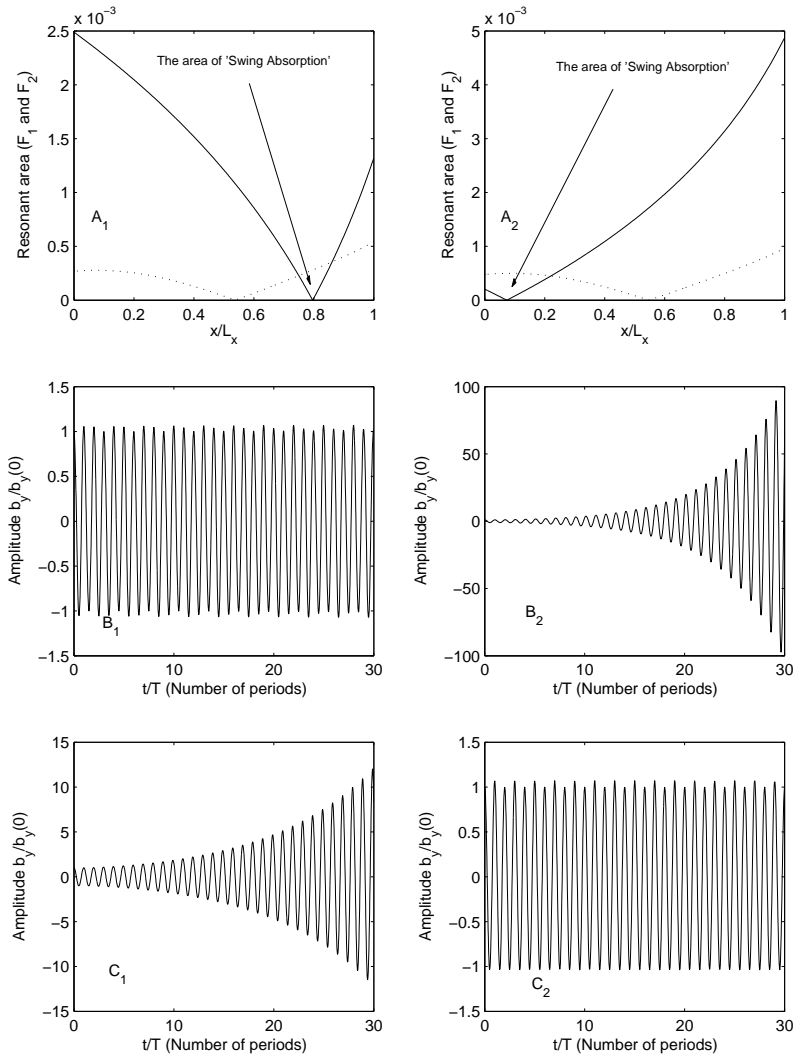}
\caption{The swing absorption of the fast mode $n=0$ by the
standing Alfv\'en modes $m=3$ and 4. }\label{fapb1}
\end{figure*}

For the configuration described by the equilibrium profiles
(\ref{pprof}) - (\ref{bprof}) the resonant condition
(\ref{rescondm}) yields the areas of swing absorption (located
along the $x$ axis) as the solutions of the following equation:
\begin{equation}
\omega _{n}C\left( p_{00}+p_{01}\frac{x_{n,m}}{L_x}\right)
^{\frac{%
1}{2\gamma }}-\frac{k_{m}}{\sqrt{\pi}}\sqrt{
h_{00}+h_{01}\frac{x_{n,m}}{L_x}} =0.
\end{equation}
In Figure~\ref{fapb1} (panels A$_1$ and A$_2$) we plot
\begin{equation}\label{f1}
F_1=\left |\frac{\omega _n}{2}-k_m V_A \right |
\end{equation}
(solid line) and
\begin{equation}\label{f2}
F_2=\left | \frac{\delta (x)}{\omega _n} \right |
\end{equation}
(dotted line) against the normalized coordinate $x/L_x$. These
curves correspond to the zeroth-order harmonic of the fast
magnetosonic mode shown in Figure~\ref{fapa1} (panel A) and the
standing Alfv\'en mode with wave numbers $m=3$ (panel A$_1$) and
$m=4$ (panel A$_2$).

In order to examine the validity of the approximations we made
during the analysis of the governing equation (\ref{eqalvin}), we
performed a direct numerical solution of the set of equations
(\ref{alfnon1}) - (\ref{alfnon2}) and obtained the following
results. The Alfv\'en mode $m=3$ is amplified effectively close to
the resonant point $x/L_x \approx x_{0,3}/L_x =0.7967$. This is
shown on panel~C$_1$ of Figure~\ref{fapb1}. Far from this resonant
point, the swing interaction is weaker, as at the point
$x/L_x\approx x_{0,4}/L_x=0.0767$ (panel B$_1$). For the Alfv\'en
mode $m=4$ we have the opposite picture: the area of {\lq swing
absorption\rq} is situated around the point $x_{0,4}$ (see panel
B$_2$, Figure~\ref{fapb1}) and the rate of interaction between
modes decreases far from this area, as at point $x_{0,3}$ (panel
C$_2$, Figure~\ref{fapb1}). In these calculations we took $\alpha
= 0.015$.

Similar results are obtained for the fast magnetosonic mode shown
in panel~B of Figure~\ref{fapa1}, corresponding to $\alpha= 0.03$.
In this case, the fast magnetosonic mode effectively amplifies
four different spatial harmonics of Alfv\'en modes, viz.\ $m=8,
10$, $11$ and $12$ (for details see Shergelashvili et al. 2004). A
similar analysis can be performed for the case of any other
equilibrium configuration and corresponding harmonics of the
standing fast magnetosonic modes.

\section{Discussion and conclusion}

The most important characteristic of swing absorption is that the
velocity polarization of the amplified Alfv\'en wave is strictly
perpendicular to the velocity polarization (and propagation
direction) of fast magnetosonic waves. This is due to the
parametric nature of the interaction. For comparison, the
well-known resonant absorption of a fast magnetosonic wave can
take place only when it does not propagate strictly perpendicular
to the magnetic flux surfaces and the plane of the Alfv\'en wave
polarization. In other words, the energy in fast magnetosonic
waves propagating strictly perpendicular (i.e.\ $k_\bot=0$) to the
magnetic flux surfaces cannot be resonantly {\lq absorbed\rq} by
Alfv\'en waves with the same frequency polarized in the
perpendicular plane. This is because the mechanism of resonant
absorption is analogous to the mechanical pendulum undergoing the
direct action of an external periodic force. This force may
resonantly amplify only those oscillations that at least partly
lie in the plane of force. On the contrary, the external periodic
force acting parametrically on the pendulum length may amplify the
pendulum oscillation in any plane. A similar process occurs when
the fast magnetosonic wave propagates across the unperturbed
magnetic field. It causes a periodical variation of the local
Alfv\'en speed and thus affects the propagation properties of the
Alfv\'en waves. As a result, those particular harmonics of the
Alfv\'en waves that satisfy the resonant conditions grow
exponentially in time. These resonant harmonics are polarized
perpendicular to the fast magnetosonic waves and have half the
frequency of these waves. Hence, for standing fast magnetosonic
waves with frequency $\omega_n$, the resonant Alfv\'en waves have
frequency ${\sim}\omega_n/2$.

In a homogeneous medium all resonant harmonics have the same
wavelengths (see Paper~I). Therefore, once a given harmonic of the
fast and Alfv\'en modes satisfies the appropriate resonant
conditions (Eqs. (23) and (25) in Paper~I), then these conditions
are met within the entire medium. Thus, in a homogeneous medium
the region where fast modes effectively interact with the
corresponding Alfv\'en waves is not localized, but instead covers
the entire system. However, when the equilibrium is inhomogeneous
across the applied magnetic field, the wavelengths of the resonant
harmonics depend on the local Alfv\'en speed. When the medium is
bounded along the unperturbed magnetic field (i.e.\ along the $z$
axis), the resonant harmonics of the standing Alfv\'en waves
(whose wavelengths satisfy condition (\ref{rescondm}) for the
onset of a standing pattern) will have stronger growth rates. This
means that the {\lq absorption\rq} of fast waves will be stronger
at particular locations across the magnetic field. In the previous
section we showed numerical solutions of standing fast
magnetosonic modes for a polytropic equilibrium ($p_0 \sim \rho _0
^\gamma$) in which the thermal pressure and magnetic pressure are
linear functions of $x$. Further, we performed a numerical
simulation of the energy transfer from fast magnetosonic waves
into Alfv\'en waves at the resonant locations, i.e.\ the regions
of {\it swing absorption}.

The mechanism of swing absorption can be of importance in a
variety of astrophysical situations.

\section*{Acknowledgments}

This work has been developed in the framework of the pre-doctoral
program of B.M.S. at the CPA, K.U.Leuven (scholarship OE/02/20).
The work of T.V.Z. was supported by the NATO Reintegration Grant
FEL.RIG 980755 and the grant of the Georgian Academy of Sciences.
These results were obtained in the framework of the projects
OT/02/57 (K.U.Leuven) and 14815/00/NL/SFe(IC) (ESA Prodex 6).

\end{document}